\documentclass[12pt]{article}

\usepackage{amsmath,amssymb,amsfonts}
\usepackage{physics}
\usepackage{bm}
\usepackage{geometry}
\title{Closed-loop Structure of Quantum Probabilities from Unitarity}
\author{M. J. Rave}

\date{}

\begin{document}

\maketitle

\begin{center}
Department of Chemistry and Physics \\
Western Carolina University \\
Cullowhee, North Carolina 28723
\end{center}

\begin{abstract}
In \cite{Rave2008} and \cite{Rave2011} it was proposed that \textit{closed loops} should be treated as fundamental quantum entities, and such loops were presented in a quasi-probability framework.
We demonstrate that the closed-loop decomposition of quantum probabilities is a \textit{direct consequence of unitarity}, and that Bargmann invariants arise naturally as the phase-invariant quantities associated with these loops, rather than being introduced independently.
This identifies interference not as mysterious cross terms, but as contributions from distinct classes of closed loops weighted by their associated Bargmann phases.
Additionally, the Born rule is seen to reflect the fundamental quadratic structure arising from the product of forward and reverse amplitudes, which together define such loops.
\end{abstract}

\section{Introduction}

Interference lies at the ``heart of quantum mechanics'' \cite{Feynman1965}, yet its probabilistic structure remains conceptually subtle.  
In particular, transition probabilities are not strictly additive under the combination of alternatives, but instead involve additional cross terms whose origins are not transparent.  
Additionally, while the Born rule provides a remarkably successful prescription for computing probabilities, its quadratic form is typically introduced as a \textit{postulate} whose underlying origin is not immediately evident.  
This suggests that a deeper understanding of quantum probabilities may be required.

Berry showed \cite{Berry1984} that physically meaningful phase-invariant quantities can arise independently of operator eigenvalues, revealing a nontrivial geometric structure underlying quantum evolution. The existence of such quantities suggests that geometric, or more generally holonomy-based, structures may play a fundamental role in quantum mechanics. In this context, Bargmann invariants \cite{Bargmann1964,MukundaSimon1993} provide a natural generalization: these are phase-invariant cyclic products of amplitudes associated with closed sequences of states, and encode geometric phase information in a discrete setting. 

In standard treatments, Bargmann invariants are typically introduced as mathematical tools for analyzing geometric phases, rather than as intrinsic quantities governing the probabilistic structure of quantum mechanics.
In \cite{Rave2008}, it was proposed that such closed-loop quantities could be treated as fundamental quantum entities.  In this framework, probabilities could be interpreted as sums over quasi-probabilities associated with loops.

Later, \cite{Rave2011} emphasized the forward/return structure of such loops, noting that for $N$ states there are $N^2$ such closed-loop combinations.  This was taken to suggest a possible connection to the $N^2$ dependence of the Born rule.

This prior work had its limitations.  The loop framework described was a heuristic approach: it was postulated, rather than derived within quantum mechanics.  And the specific role of Bargmann invariants was not structurally grounded.

In this paper, we demonstrate that the closed-loop decomposition of quantum probabilities is a direct consequence of unitarity (i.e.\ $U U^\dagger = U^\dagger U = I$).  We also show that probabilities necessarily involve both forward and reverse amplitudes, and that algebraic expansion produces closed loop sequences \textit{automatically}. In this formulation, Bargmann invariants emerge intrinsically (not introduced ad hoc) and naturally appear as the phase-invariant factors in each loop.

\section{Closed Loops and Phase-Invariant Products}

We adopt the convention that amplitudes are denoted by
\begin{equation}
\phi_{ab} \equiv \langle a|U|b\rangle,
\end{equation}
where $a$ and $b$ are arbitrary states and $U$ is a unitary operator.
Amplitudes (and later, products of amplitudes) are interpreted as sequences of states read from
right to left. Thus $\phi_{ab}$ represents a directed transition
$b \to a$.
In the absence of explicit evolution (i.e. $U = I$), this reduces
to $\phi_{ab} = \langle a|b\rangle$.
We define a closed loop sequence as
\begin{equation}
S = (s_1 \to s_2 \to \cdots \to s_N \to s_1).
\end{equation}
To each such loop we associate the product
\begin{equation}
\Gamma(S)
=
\prod_{k=1}^{N}
\phi_{s_{k+1},\, s_k},
\qquad
s_{N+1} \equiv s_1.
\end{equation}

In general, $\Gamma(S)$ is a complex number whose magnitude and phase are determined by the amplitudes along the loop.
Note that $\Gamma(S)$ is invariant under arbitrary phase choices of the participating states, as are Bargmann invariants and the Berry phase.

Define the phase associated with a loop $S$ by
\begin{equation}
\gamma(S) = -\mathrm{Im}\,\ln \Gamma(S).
\end{equation}
The phase $\gamma(S)$ is, like $\Gamma(S)$, invariant under arbitrary phase choices.

In \cite{Rave2008}, transition probabilities were proposed to take the form
\begin{equation}
P_{fi} = \sum_{S \ni (i,f)} \Gamma(S),
\end{equation}
where the sum is over all closed loops containing $i$ and $f$.  In this paper we prove, and expand upon, this result.

\section{Closed-loop Structure from Unitarity}
Having described closed-loop sequences in \textit{abstract} terms, we now show how they arise directly from the unitary structure underlying quantum probabilities.
Note that in \cite{Rave2008}, amplitudes were represented by inner products of the form
$\phi_{fi} = \langle f|i\rangle$ (without any intermediate operator), emphasizing a formulation in terms of overlaps.
Here we introduce an explicit unitary operator $U$ so that $\phi_{fi} = \langle f|U|i\rangle$ defines a general transition amplitude,
with the convention that amplitudes are read from right to left, so that $\phi_{fi}$ corresponds to the transition $|i\rangle \to |f\rangle$.
With this definition, the closed-loop structure emerges directly from the quadratic form $|\langle f|U|i\rangle|^2$, rather than being postulated.

It should be emphasized that this ordering reflects the structure of amplitude composition, and does not imply \textit{temporal} evolution.
The unitary operator $U$ \textit{may} represent time evolution (e.g. $U = e^{-iHt/\hbar}$), but the derivation that follows depends only on its unitarity, not on any
specific dynamical interpretation or choice of Hamiltonian.

\subsection{Statement of Theorem}

Let $U$ be a unitary operator on a finite-dimensional Hilbert space, and define transition amplitudes by
\begin{equation}
\phi_{fi} \equiv \langle f|U|i\rangle.
\end{equation}
Then the transition probability
\begin{equation}
P_{fi} = |\phi_{fi}|^2
\end{equation}
admits the exact decomposition
\begin{equation}
\boxed{
P_{fi}
=
\sum_{n,m}
\phi_{i m}\,\phi_{m f}\,\phi_{f n}\,\phi_{n i}
}
\end{equation}
where $n$ and $m$ label independent intermediate states arising from two resolutions of the identity.  Each term in this sum is a phase-invariant product of amplitudes associated with a closed sequence of states
\begin{equation}
S = (i \to n \to f \to m \to i),
\end{equation}
so that the transition probability is expressed as a sum over closed loops passing through $i$ and $f$.  As will be shown, this construction extends straightforwardly to closed loops of arbitrary length.

\medskip
\noindent\textit{In words:} The quadratic structure of quantum probabilities gives rise to a natural pairing of forward and reverse amplitudes, and this pairing generates closed-loop sequences in state space.
The resulting loop structure emerges directly from unitarity, rather than being imposed independently, and suggests that such closed loops may provide a more fundamental perspective on the origin of the quadratic form itself.

\subsection{Proof}

We begin with
\begin{equation}
P_{fi} = |\langle f|U|i\rangle|^2,
\end{equation}
and insert a resolution of the identity,
\begin{equation}
I = \sum_n |n\rangle \langle n|,
\end{equation}
where $\{|n\rangle\}$ is any orthonormal basis,
to obtain
\begin{equation}
\phi_{fi}
=
\langle f|U|i\rangle
=
\sum_n \langle f|U|n\rangle \langle n|i\rangle
=
\sum_n \phi_{fn}\,\phi_{ni}.
\end{equation}

Thus
\begin{equation}
P_{fi}
=
\left( \sum_n \phi_{fn}\,\phi_{ni} \right)
\left( \sum_m \phi_{fm}\,\phi_{mi} \right)^*
=
\sum_{n,m}
\phi_{fn}\,\phi_{ni}\,
\phi_{mf}\,\phi_{im},
\end{equation}
where we have used $\phi_{fm}^* = \phi_{mf}$ and $\phi_{mi}^* = \phi_{im}$.

Reordering the factors yields
\begin{equation}
P_{fi}
=
\sum_{n,m}
\phi_{im}\,\phi_{mf}\,\phi_{fn}\,\phi_{ni}.
\end{equation}

Each term therefore has the form $\Gamma(S) = \phi_{im}\,\phi_{mf}\,\phi_{fn}\,\phi_{ni}$, corresponding to the closed sequence $S = (i \to n \to f \to m \to i)$, so that
\begin{equation}
P_{fi} = \sum_{S \ni (i,f)} \Gamma(S),
\end{equation}
as in \cite{Rave2008}.

This construction extends naturally by inserting additional resolutions of the
identity. In this way, $P_{fi}$ may be expressed as a sum over closed sequences
of arbitrary length,
\begin{equation}
S = (i \to s_1 \to s_2 \to \cdots \to s_L \to i),
\end{equation}
with each contribution given by a corresponding product
$\Gamma(S)$.  At this stage, the probability is expressed as a sum over individual closed loops,
each contributing a generally complex quantity $\Gamma(S)$.

Since the total probability must be real, it is natural to group each loop with its reversed
sequence.  Accordingly, for each loop $S$ we associate a reversed sequence $S^{-1}$,
obtained by reversing the ordering of the states. By construction, the corresponding
amplitudes satisfy
\begin{equation}
\Gamma(S^{-1}) = \Gamma(S)^*,
\end{equation}
so that the contribution of each pair $(S,S^{-1})$ is real.

Writing $\Gamma(S) = |\Gamma(S)| e^{i\gamma(S)}$, with
$\gamma(S) = -\mathrm{Im}\ln \Gamma(S)$, one finds
\begin{equation}
\Gamma(S) + \Gamma(S^{-1})
=
2|\Gamma(S)|\cos(\gamma(S)).
\end{equation}
Summing over such pairs yields
\begin{equation}
P_{fi}
=
\sum_{S}
2|\Gamma(S)|\cos(\gamma(S)).
\end{equation}

Though it bears a superficial resemblance to
path-integral formulations, the present approach is distinct in that the
decomposition arises entirely from the unitary structure of the Hilbert
space and operates at the level of probabilities rather than amplitudes.
The closed loops that appear are therefore not dynamical trajectories,
but algebraic consequences of the quadratic form.

\subsection{Interpretation}

The above decomposition admits a simple interpretation. The probability
$P_{fi}$ arises from pairing a forward amplitude with an independent
return amplitude, producing closed sequences of the form
\begin{equation}
i \to n \to f \to m \to i.
\end{equation}

Since the forward and return processes each involve a sum over $N$
intermediate states, their combination yields $N^2$ distinct
forward–return pairings. Each such pairing defines one such closed loop
in state space, and the transition probability is obtained by summing
over all such loops. This provides a structural origin for the quadratic
form of the Born rule: it reflects the combinatorics of forward–return
pairings.

It is important to emphasize that no assumption has been made regarding time evolution. While the decomposition may be visualized in terms of forward and backward evolution, the indexing of states is more general.
In particular, the closed sequences are analogous to parameter-space loops underlying Berry phases, suggesting that the loop structure reflects Hilbert space geometry rather than temporal dynamics alone.

In \cite{Rave2011}, additional consistency conditions were required to constrain the allowed closed sequences. This is because, when probabilities were expressed in terms of transitions $|i\rangle \to |f\rangle$, cross terms appeared that violated classical additivity.
In the present formulation, however, this non-additive structure emerges directly from the algebra of unitary amplitudes, without the need for additional assumptions.  In the closed-loop representation $|i\rangle \to \cdots \to |i\rangle$,
each loop contributes a single term $\Gamma(S)$, and probabilities arise as sums over such contributions.

\subsection{Implications of the Loop Decomposition}

The representation of transition probabilities as
\begin{equation}
P_{fi} = \sum_S 2|\Gamma(S)|\cos(\gamma(S))
\end{equation}
has a number of immediate consequences.

First, each term in the probability is individually phase invariant. The magnitude $|\Gamma(S)|$ represents the \textit{weight} associated with a given loop, while the phase $\gamma(S)$ determines whether that term enhances or suppresses the total probability.
In this way, constructive and destructive interference arise directly from the \textit{sign} of $\cos(\gamma(S))$. Interference is therefore not an additional feature imposed ad hoc upon probabilities, but is a direct consequence of the phase structure associated with closed loops.

Second, this equation suggests a natural mechanism underlying decoherence.  When interactions with an environment render different alternatives distinguishable, the associated loop phases lose coherence.
In effect, the phases $\gamma(S)$ vary in an uncontrolled manner, and the corresponding cosine factors average to zero. The suppression of interference may thus be understood as a form of phase randomization over loops.
This interpretation will be developed in more detail in the following subsection.

Finally, the dependence of each term on the phase $\gamma(S)$ highlights the \textit{geometric} content of the decomposition. Since these phases are closely related to Bargmann invariants and discrete Berry phases,
the above expression makes explicit that quantum interference is governed by geometric phases associated with closed sequences in Hilbert space. Our loop decomposition therefore provides a framework in which probability, interference,
and geometric phase are unified at the structural level.

\subsection{Decoherence}

The phase-randomization picture described above finds a direct realization in decoherence \cite{Zurek2003}. In this framework, the suppression of non-self-retracing loops reflects the emergence of distinguishability between alternative histories.
Within this setting, the disappearance of interference in macroscopic regimes may be understood as a suppression of loop contributions involving distinct alternatives. When intermediate states encode path information,
the corresponding products $\Gamma(S)$ vanish or become negligible, leaving only self-retracing loops and restoring classical additivity of probabilities.

While, in general, all closed loops consistent with the algebraic expansion of $P_{fi}$ contribute to the probability, the presence or absence of particular loops depends on whether intermediate states retain information about the path taken.
When such path dependence is present, mixed (non-self-retracing) loops involve distinguishable alternatives and their contributions are suppressed, so that only self-retracing loops remain. In this light,
the additional consistency conditions introduced in \cite{Rave2011} are seen to reflect the suppression of non-self-retracing loops due to path distinguishability, rather than constituting independent assumptions.

The removal of such non-self-retracing loops corresponds to the disappearance of interference terms, providing a structural account of decoherence as the elimination of a subset of closed-loop contributions.

\section{Conclusion}

We have shown that the closed-loop decomposition of quantum probabilities arises directly from the unitary structure of amplitudes. In this formulation, probabilities are expressed as sums over phase-invariant loop contributions,
with each loop characterized by an associated phases $\Gamma(S)$ and $\gamma(S)$.

This perspective leads to several conceptual reinterpretations.  Interference may be viewed as being governed by loop holonomies, with constructive and destructive contributions determined by the associated phases. In particular,
interference may be understood as arising from non-self-retracing loops, while the disappearance of interference corresponds to the \textit{suppression} of such loops. The Born rule itself is seen to reflect the underlying quadratic structure of amplitudes,
arising from the pairing of forward and return processes.  Closed loops, rather than transitions, become the natural probabilistic objects of quantum mechanics.

The present formulation suggests a number of directions for further investigation. It would be of interest to explore more systematically the role of loop phases in semiclassical regimes, and to examine whether stationary-phase arguments can be formulated at the level of loop contributions.
The connection between loop structure and decoherence may also be developed further, particularly in relation to environment-induced phase randomization.
Additionally, the relationship between the present framework and other formulations of quantum theory (such as path integrals, consistent histories, and geometric phase approaches) merits deeper study.

Taken together, these results suggest that quantum probability is fundamentally a geometric phenomenon, determined by the structure and phases of closed loops in Hilbert space.

\end{document}